\documentclass[12pt]{iopart}

\usepackage{dcolumn}
\usepackage{bm}
\usepackage{verbatim}       

\usepackage[dvips]{graphicx}
\usepackage{amsthm}
\usepackage{amssymb}
\usepackage{bm}
\usepackage{amstext}
\usepackage{psfrag}

\expandafter\let\csname equation*\endcsname\relax

\expandafter\let\csname endequation*\endcsname\relax

\usepackage{amsmath}
\usepackage{amsfonts}
\usepackage{mathrsfs}
\usepackage{cite}

\newcommand{\p}{\partial}

\newcommand{\dd}{{\rm d}}

\newcommand{\bd}{\begin{definition}}                
\newcommand{\ed}{\end{definition}}                  
\newcommand{\bc}{\begin{corollary}}                 
\newcommand{\ec}{\end{corollary}}                   
\newcommand{\bl}{\begin{lemma}}                     
\newcommand{\el}{\end{lemma}}                       
\newcommand{\bp}{\begin{proposition}}            
\newcommand{\ep}{\end{proposition}}                
\newcommand{\bere}{\begin{remark}}                  
\newcommand{\ere}{\end{remark}}                     

\newcommand{\bt}{\begin{theorem}}
\newcommand{\et}{\end{theorem}}

\newcommand{\bit}{\begin{itemize}}
\newcommand{\eit}{\end{itemize}}
\newtheorem{theorem}{Theorem}[section]
\newtheorem{corollary}[theorem]{Corollary}
\newtheorem{lemma}[theorem]{Lemma}
\newtheorem{proposition}[theorem]{Proposition}
\theoremstyle{definition}
\newtheorem{definition}[theorem]{Definition}
\theoremstyle{remark}
\newtheorem{remark}[theorem]{Remark}

\begin{document}

\title{The conformal transformation of the night sky}

\author{E Minguzzi}
\address{Dipartimento di Matematica e Informatica ``U. Dini'', Universit\`a
degli Studi di Firenze, Via S. Marta 3,  I-50139 Firenze, Italy.}

\ead{ettore.minguzzi@unifi.it}


\date{}

\begin{abstract}
\noindent
 We give a simple differential geometric proof of the conformal transformation of the night sky under change of observer. The proof does not use the four dimensionality of spacetime or   spinor methods. Furthermore, it really shows that the result does not depend on Lorentz transformations.
 This approach, by giving a transparent covariant expression to the conformal factor, shows that in most situations it is possible to define a {\em thermal sky metric} independent of the observer.
\end{abstract}

\section{Introduction}
To any observer the night sky appears as a  distribution of stars and constellations which could be mapped over a sphere $S^2$ called the {\em  sky}.
This notion, converted into an appropriate geometrical object, can be found  in foundational studies of relativity \cite{komar65}, as a key concept for expressing the relativity principle, or in studies of the causal structure of spacetime \cite{low89,low06,bautista14,natario04,chernov10}.

Two observers at the same spacetime event $x$ but relatively boosted will perceive a different night sky due to the phenomenon of stellar aberration.
In the late fifties  papers by Terrell \cite{terrell59} and Penrose \cite{penrose59} showed that the map between  skies is conformal.
One might assume that the result  depends on the Lorentz group, and should be derived making use of appropriate Lorentz transformations.
However, in this paper we shall show that it is much more robust and general, for it does not rely on symmetries of the observer space (indicatrix). In relativity theory this space is $\mathbb{H}^3$ and so it is homogeneous. Furthermore, we shall show that in many cases it is possible to introduce {\em an invariant metric} on $S^2$, not just a conformal one. In order to achive this result one should also consider the frequency of photons and not just their direction. In other words one should consider the full sky picture, not just points and shapes but also {\em color}.

In the remainder of this introduction let us recall the usual methods of proof. We use conventions according to which $c=1$ and the signature of the spacetime metric $g$ is $(-,+,\cdots,+)$.  The most straightforward argument is based on the Lorentz transformation and goes as follows \cite{terrell59,peres87}.

Let two inertial observer $K$ and $K'$ be boosted and let coordinates be chosen so that the boost is in the $z$ direction for both observers, so that the velocity of $K'$ with respect to $K$ is $(0,0,v)$. Let $u$, $u'$ be the velocities of any other particle as measured by the observers. The Lorentz transformation for velocities is
\begin{align*}
u'_z&=\frac{u_z-v}{1-v u_z}, \quad u'_{x}= \frac{u_{x}}{\gamma (1-v u_z)}, \quad u'_{y}= \frac{u_{y}}{\gamma (1-v u_z)}
\end{align*}
Let $(\theta,\phi)$ and $(\theta',\phi')$ be polar coordinates for $K$ and $K'$, respectively. If the particle is really a photon coming from direction $(\theta,\phi)$ for $K$ then $u_z=-\cos \theta$,
and analogously for $K'$ thus since $u_x/u_y=u'_x/u'_y$
\begin{align*}
\cos \theta'&=\frac{v+\cos \theta}{1+v \cos \theta}, \\
\phi'&=\phi.
\end{align*}
This is the map $\varphi\colon S^2 \to S^2$ between skies. Differentiation of the former equation gives $\dd \theta' /\sin \theta'=\dd \theta/\sin \theta$, thus denoting the canonical metric of $S^2$ with $\dd \Omega^2= \dd \theta^2+\sin^2 \theta \dd \phi^2$, the map between skies satisfies
\begin{equation} \label{mis}
(\dd \Omega')^2=\left(\frac{\sin \theta'}{\sin \theta}\right)^2 \dd \Omega^2 ,
\end{equation}
namely, it is conformal. Penrose goes on to observe that the stereographic projection of $S^2$ on the complex plane $z=0$ is conformal.  The map between skies is a conformal bijection of the extended complex space (Riemann sphere), and any conformal map of this type is really a fractional linear (M\"obius) transformation,\footnote{Any  conformal bijection between open sets of the complex plane is biholomorphic but the  {\em local} conformal property of the map by itself does not guarantee the preservation of circles. The general lack of rigidity is also suggested by the Riemann mapping theorem according to which any non-empty open simply connected proper subset of the complex plane admits a conformal bijection to the unit disk.} and so sends circles into circles \cite{penrose59}.
The most general transformation is a composition of a rotation, a  boost and a rotation, where rotations are trivially conformal and send circles into circles, so the just proved results  hold for general transformations as well. 

Given a uniformly moving sphere  on Minkowski spacetime there is an observer  at the observation event for which the sphere is at rest and which therefore sees it with a circular profile. As a consequence, for any observer  the profile of the sphere looks circular.

The conformality of the sky map  $\varphi\colon S^{n-1} \to S^{n-1}$ holds also for spacetimes of dimension $n+1$. The proof remains the same, the angle $\phi$ would be dropped (for $n=2$) or replaced by more angles $\phi_i, i=1,c\dots n-2$, left unchanged by the boost. Also by Liouville's theorem it is still true that  $S^{n-2}$ spheres are sent to spheres.



Penrose gave also a rather different proof of the conformality of the map based on spinor geometry \cite{penrose59,penrose84,naber92}. While his approach gives other insights into this map, it  works just in four spacetime dimensions.

\section{A differential geometric argument}

Let  $(M,g)$  be a spacetime, namely a time oriented Lorentzian manifold, and let us place our considerations on just a tangent space $V:=T_xM$, where $x\in M$ is the event of observation.

Let $\{ x^\mu\}$ be local coordinates on $M$ and let $\{x^\mu, y^\mu\}$ be induced coordinates on $TM$. Let $\Omega_x\subset T_xM$ be the future timelike cone at $x$ and let $\mathscr{N}_x=\p \Omega_x$ be its boundary (the closure operator is in the slit tangent bundle $TM\backslash 0$) namely the light cone, i.e.\ the set of future lightlike vectors. Observe that $\Omega_x$ is an  open convex sharp cone for every $x$ (sharp means that $\Omega_x$ does not contain lines).

The {\em observer space} at $x$  is
\[
\mathscr{I}_x:=\{y\in \Omega_x: g(y,y)=-1\}.
\]
It is also called {\em velocity space} or {\em indicatrix}. Each point represents the covariant velocity of a massive particle/observer.


For every $x\in M$, $(T_xM ,g) $ can be regarded as a flat Lorentzian manifold.  The manifold $\mathscr{N}_x$ on the spacetime $(T_xM , g)$ is lightlike since for every $y\in \mathscr{N}_x$ the  vector $y \in T_y \mathscr{N}_x$ is lightlike, $g(y,y)=0$.

For $u \in \mathscr{I}_x$, the hyperplane on $T_xM$ tangent to the indicatrix at $u$ is
\[
 \mathscr{P}_u
 = \{y\in T_xM: g(u,y-u)=0\} ,
\]
thus it is orthogonal to the timelike vector $u\in T_u\Omega_x$.

Given an observer $u\in \mathscr{I}_x$  we define the {\em sky} or {\em celestial sphere} of $u$ as follows
\[
\mathscr{S}_u=\mathscr{N}_x\cap \{y\in T_xM: \ g(u, y-u)=0\}=\{y \in \mathscr{N}_x: \ g(u,y)=-1 \} .
\]
Actually the sky is more properly $-\mathscr{S}_u$ (see \cite{penrose84}), however, we shall work with the above definition since it will allow us to place all the geometrical considerations in the  future cone $\Omega_x$.

The sky can be naturally endowed with a metric which is just the metric induced from $g$. Let $v,w\in T_y \mathscr{S}_u$ and define
\begin{equation}
\gamma(y)(v,w):=g(v,w).
\end{equation}
This is a Riemannian metric on $\mathscr{S}_{u}$ since, as mentioned, $\mathscr{N}_x$ is lightlike \cite{hawking73}.


Given an observer $u$ we introduce {\em observer coordinates} $\{y^\mu\}$ on $T_xM$ with the following condition $y^0(u)=1$, $y^i(u)=0$, and $g=-(\dd y^0)^2+\sum_i (\dd y^i)^2$. In practice the $y^0$ axis is oriented with $u$ while the space axes are parallel to the hyperplane tangent to the indicatrix at $u$ and are oriented so as to diagonalize the metric $g$. With these coordinates the metric $g$ is just the Minkowski metric in its canonical form and $\mathscr{S}_u$ is  the set $\{y\colon y^0=1, \sum_i (y^i)^2=1\}$, thus the sky is a sphere $S^{n-1}$  endowed with the canonical metric.

\begin{figure}[ht]
\begin{center}
 \includegraphics[width=6cm]{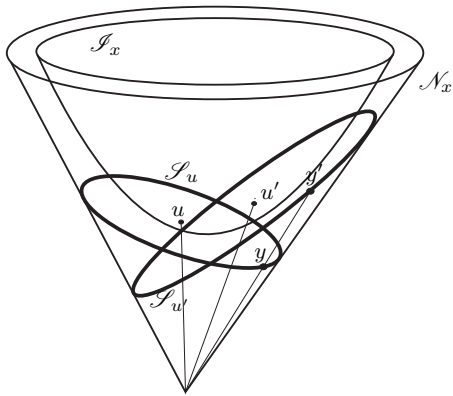}
 \caption{The cone $\Omega_x$ and the geometrical construction related to the conformal map of skies} \label{squ}
\end{center}
\end{figure}

Let us consider two observers $u, u'\in \mathscr{I}_x$ and their skies $\mathscr{S}_{u}$ and $\mathscr{S}_{u'}$. Let a light ray reach both observers and let a generator of $\mathscr{N}_x$ fix the direction of the photon velocity. This direction determines points $y,y'\in \mathscr{N}_x$ on the observer's skies (Fig.\ \ref{squ}).
 More precisely the  projection on the projective space of $T_xM$ determines a  bijection $\varphi\colon \mathscr{S}_{u}\to \mathscr{S}_{u'}$ between the skies where each point is sent to the point with the same projection.

 Now, there is some $s>0$ such that $y'=sy$. Furthermore, defined $v'=\varphi_* v$, $w'= \varphi_* w$,  we have, since $v$ and $v'$ have the same projection, $v'=s v+ a y$, $a\in \mathbb{R}$, and analogously $w'=s w+b y$, with $b\in \mathbb{R}$.
Next observe that
\[
{g(v',w')}=s^2 \left [\frac{g(y,u)}{g(y',u')}\right]^2 g(v,w)
\]
since both numerator and denominator of the fraction equal -1, thus
\begin{equation} \label{odg}
\gamma'(v',w')= \left[\frac{g(y,u)}{g(y,u')}\right]^2 \gamma(v,w).
\end{equation}
This equation proves that the map $\varphi$ is conformal. It can be observed that the ratio
\begin{equation} \label{moa}
D:=\frac{g(y,u)}{g(y,u')}
\end{equation}
and hence the conformal factor is invariant under replacements $y \to \lambda y$. Equation (\ref{odg}) provides a transparent covariant expression for the conformal factor. It depends on the direction of sight $y\in \mathscr{N}_x$ and on the two observers $u,u'\in \mathscr{I}_x$ whose observations are being related.

An important fact which does not seem to have been sufficently stressed is the following \cite{terrell59}
 \begin{proposition}
The conformal factor relating the sky metrics is $D^2$, where $D$ is the Doppler factor, namely the ratio of the photon frequencies as observed by $u$ and $u'$.
\end{proposition}
This fact follows at once since for some   $\lambda$, $p=g(\lambda y,\cdot)$ is the momenta of the photon.\footnote{In Eq.\ (\ref{mis}) we can write $\sin \theta'/\sin \theta={\sqrt{1-v^2}}/{(1+v \cos \theta)}$. In the coordinates of observer $K$, $u=(1,0,0,0)$, $u'=\frac{1}{\sqrt{1-v^2}}(1,0,0,v)$, and the contravariant momenta of the photon is $h \nu (1,-\sin \theta, \cos \phi,-\sin \theta \sin \phi, -\cos \theta)$, so one can easily check that the conformal factor is the squared Doppler factor.}
Since the relationship between the area forms on the skies induced by the conformal map is
\begin{equation}
\varphi^* \dd v'=\left[\frac{g(y,u)}{g(y,u')}\right]^{n-1} \dd v
\end{equation}
we conclude (under the assumption that from each direction there arrive photons of just one direction-dependent frequency)

\begin{theorem}
In the observation of the night sky the product between the area of the observed patch of sky and the frequency of photons at power $n-1$ is independent of the observer.
\end{theorem}

Thus if we boost towards the Big Dipper the stars frequency  will increase but the constellation will shrink since the product of its angular area times the frequency squared must be constant.

Another suggestive interpretation of (\ref{odg}) written in the form
\begin{equation} \label{lod}
\langle p, u'\rangle^2 \gamma'= \langle p, u\rangle^2 \gamma
\end{equation}
 is this:
 \begin{quote}
 {\em the sky has an intrinsic invariant metric which is revealed once the wavelength of the observed photons is used as unit angular scale}.
 \end{quote}
 Stated in another way:
  \begin{quote}
  {\em if we measure the sky with the  natural lengths provided by the sky itself, it shows the same geometry to all observers}.
\end{quote}
These interpretations are valid provided the related observers have means to select the photons to be measured from the whole spectra they receive from each direction.
For instance, let us consider a polyhedra moving on Minkowski spacetime and let us suppose that in the rest frame every point of the surface emits light in any direction at the same frequency. Then at the observation event any two observers taking snapshots would get different conformally related images.
 By taking into account the color of their own picture they can, by means of the metric (\ref{lod}), reach a value for the (colored) angular distance between vertices which is  really independent of the observer. All that without communicating among themselves.
They can even recover the shape that would be seen by that one observer who, being at rest with respect to the polyhedra, gets the whole picture  in the same uniform color.

Similarly, one can consider the situation in which the object emits a black body radiation at temperature $T_0$. Let the observers agree to focus on any given direction to that spectral band which corresponds to the maximal intensity in the frequency domain (these frequencies differ among the frames but select the same photons). By Wien's law the  {\em thermal sky metric}
\[
\tau:=T^2(\Omega) \dd \Omega^2
\] is independent of the observer, where $T$ is the observed temperature in direction $\Omega$.

Finally, let us   consider the cosmic microwave background radiation (CMB) and suppose that the two observers have agreed to retain, for each direction, just that portion of waveband which corresponds to the maximal intensity in the frequency domain. As a result they are observing the same photons and so the thermal sky metric $T^2(\Omega) \dd \Omega^2$ is again independent of the observer.



Perhaps, the best way to geometrically understand this result is by working on the slit contangent bundle $T^*M\backslash 0$. We can repeat similar constructions: define the polar cone $\Omega_x^*$,  its boundary $\mathscr{N}_x^*$, and the metric $g^{-1}$. Then given a section $s\colon P\mathscr{N}_x^*  \to \mathscr{N}_x^*$, namely given a photon momenta for every line of sight, we can define a metric on $P\mathscr{N}_x^*$  by $g^{-1}(V,W)$ where $V$ and $W$ are representatives for $v,w\in TP\mathscr{N}_x^*$. This metric coincides with (\ref{lod}), though the former clarifies the conformal connection with the usual sky metric.

\section{The Finslerian version}

In this section we consider a more general Finslerian framework.
We shall be able to repeat the previous arguments thus showing that  the conformality of the transformation of skies is a robust result which is independent of the symmetries of the observer space. This fact shows once again that Lorentz transformations are not required for the derivation of the result.


Let $\Omega\subset TM$ be a distribution of open convex sharp cones on the tangent bundle $\pi\colon TM\to M$, with vertex in the zero section.
This means that $\Omega_x$ is an  open convex sharp cone for every $x$.


A {\em Finsler Lagrangian} is a map  $\mathscr{L}\colon  \Omega \to \mathbb{R}$ which
is  positive homogeneous of degree two in the fiber coordinates
\[
\mathscr{L}(x,sy)=s^2 \mathscr{L}(x,y), \qquad \forall s>0.
  \]
  In this work we assume that the fiber dependence  is at least $C^2(\bar \Omega)$,
 that $\mathscr{L}<0$ on $\Omega$ and that $\mathscr{L}$ can be $C^2$-continuously extended setting $\mathscr{L}=0$ on $\p \Omega$. Here closure is understood in the slit tangent bundle $TM\backslash 0$.
The matrix metric is defined as the Hessian  of $\mathscr{L}$ with respect to the fibers
\[
g_{\mu \nu}(x,y)= \frac{\p^2 \mathscr{L}}{\p y^\mu \p y^\nu},
\]
and we assume that it can be continuously extended  to $\p \Omega$ preserving the Lorentzian signature.
This matrix can be used to define a metric in two different, but essentially equivalent  ways. The Finsler metric is typically defined as $g=g_{\mu \nu}(x,y) \dd x^\mu \dd x^\nu$ and is a map $g\colon \Omega \to  T^*M \otimes T^*M$.  For any given $x$ one could also use this matrix  to define a vertical metric on $\Omega_x$ as follows $g_{\mu \nu}(x,y) \dd y^\mu \dd y^\nu$. Usually the context clarifies which one is used, nevertheless, in this work we shall use the latter metric.
In index free notation the metric will be also denoted  $g_y$ to stress the dependence on the fiber coordinates.

By positive homogeneity we have
\[
\mathscr{L}=\frac{1}{2} \,g_y(y,y), \qquad g_{sy}=g_y \quad \textrm{ and} \quad\dd \mathscr{L}=g_y(y,\cdot).
\]
The usual Lorentzian-Riemannian case is obtained for $\mathscr{L}$ quadratic in the fiber variables.
 The vectors belonging to $\Omega_x$ are called {\em timelike} while those belonging to $\p \Omega_x$ are called {\em lightlike}.

%
%

The {\em observer space} is
\[
\mathscr{I}_x:=\{y\in \Omega_x: g_y(y,y)=-1\}=\mathscr{L}^{-1}(-1/2).
\]
while the light cone is
\[
\mathscr{N}_x:=\{y\in \Omega_x: g_y(y,y)=0\}=\p \Omega_x=\mathscr{L}^{-1}(0).
\]
Observe that we have just the future version of these objects unlike in Lorentzian geometry. This is so because $\mathscr{L}$ is defined on just a cone $\Omega$.

The Legendre map \cite{minguzzi13c} is $\ell:\bar \Omega \to \bar \Omega^*$, $y \mapsto g_y(y,\cdot)=\p \mathscr{L}/\p y$ where $\Omega^*$ is the polar cone.

For every $x\in M$, $\bar \Omega_x $ is really a Lorentzian manifold with boundary  $\mathscr{N}_x$ since $g_y$ is a Lorentzian metric for every $y\in \bar\Omega_x$.  The manifold $\mathscr{N}_x$ on the spacetime $(\bar \Omega_x, g_y)$ is lightlike since for every $n\in \mathscr{N}_x$ the  vector $n \in T_n \mathscr{N}_x$ is $g_n$-lightlike, $g_n(n,n)=0$.

For $u \in \mathscr{I}_x$, the hyperplane on $T_xM$ tangent to the indicatrix at $u$ is
\[
 \mathscr{P}_u=u+ \textrm{ker}\, \dd \mathscr{L}\vert_u= \{y: g_u(u,y-u)=0\}
\]
thus it is $g_u$-orthogonal to the $g_u$-timelike vector $u\in T_u\Omega_x$  (recall that $g_u(u,u)=-1$).

Given an observer $u\in \mathscr{I}_x$  we define the  sky of $u$ as follows
\[
\mathscr{S}_u=\mathscr{N}_x\cap \{y\in \bar \Omega_x: \ g_u(u, y-u)=0\}=\{y \in \mathscr{N}_x: \ g_u(u,y)=-1 \}
\]
Again the sky is more properly $-\mathscr{S}_u$. The sky is topologically $S^{n-1}$ since it is the boundary of a connected open bounded convex set \cite[Prop. 1]{minguzzi13c}

Let us show that the sky can be naturally endowed with a metric. Let $v,w\in T_y \mathscr{S}_u$ and define
\begin{equation}
\gamma(y)(v,w):=\frac{g_y(v,w)}{[g_{y}(y,u)]^2}.
\end{equation}
This is a Riemannian metric on $\mathscr{S}_{u}$ since, as mentioned, $\mathscr{N}_x$ is lightlike.

Let us consider two observers $u, u'\in \mathscr{I}_x$ and their skies $\mathscr{S}_{u}$ and $\mathscr{S}_{u'}$. Let a light ray reach the eye of both observers and let a generator of $\mathscr{N}_x$ fix the direction of the photon velocity. This direction determines points $y,y'\in \mathscr{N}_x$ on the observer's skies (Fig.\ \ref{squ}).
 More precisely the  projection on the projective space of $T_xM$ determines a  bijection $\varphi\colon \mathscr{S}_{u}\to \mathscr{S}_{u'}$ between the skies where each point is sent to the point with the same projection.

%

 Now, there is some $s>0$ such that $y'=sy$ thus $g_y=g_{y'}$. Furthermore, defined $v'=\varphi_* v$, $w'= \varphi_* w$,  we have since $v$ and $v'$ have the same projection, $v'=s v+ a y$, $a\in \mathbb{R}$, and analogously $w'=s w+b y$, with $b\in \mathbb{R}$.
Next observe that
\[
\frac{g_{y'}(v',w')}{[g_{y'}(y',u')]^2}=s^2 \left [\frac{g_y(y,u)}{g_{y'}(y',u')}\right]^2 \frac{g_{y}(v,w)}{[g_{y}(y,u)]^2}
\]
thus
\begin{equation} \label{odh}
\gamma'(y')(v',w')= \left[\frac{g_y(y,u)}{g_{y}(y,u')}\right]^2 \gamma(y)(v,w).
\end{equation}
This equation proves that the map $\varphi$ is conformal. Once again the ratio
\begin{equation} \label{mob}
D:=\frac{g_y(y,u)}{g_y(y,u')}
\end{equation}
and hence the conformal factor is invariant under replacements $y \to \lambda y$.  It gives the ratio of the
  photon frequencies as observed by $u$ and $u'$ (because for some $\lambda$, $p=g_{\lambda y}(\lambda y,\cdot)$ is the momenta of the photon).
  Thus Eq.\ (\ref{lod}) and its interpretation pass to the Finslerian case though it should be observed that the sky metric is not isometric to the canonical one of $S^{n-1}$.

\subsection{The affine metric}



%

Let us investigate the geometrical interpretation of the sky metric in the Finslerian case. Given an observer $u$ we introduce {\em observer coordinates} $\{y^\mu\}$ on $T_xM$ with the following condition $y^0(u)=1$, $y^i(u)=0$, and $g_u=-(\dd y^0)^2+\sum_i (\dd y^i)^2$. In practice the $y^0$ axis is oriented with $u$ while the space axes are parallel to the hyperplane tangent to the indicatrix at $u$ and are oriented so as to diagonalize the metric $g_u$. The observer coordinates are determined up to rotations. Next we let $v^i=y^i/y^0$. We can imagine the  $v^i$-axes as parallel to the $y^i$-axes but with origin in $u$. For $y \in \mathscr{I}_x$, the vector ${\bm v}$ gives the velocity of particle $y$ as measured by $u$. The domain of possible velocities of massive particles $D_u\subset \mathbb{R}^n$  is a convex open set which can be identified with $\Omega_x\cap \mathscr{P}_u$ so its  boundary can be identified with $\mathscr{S}_u$. Now $\mathscr{S}_u$ being a convex boundary admits an affine metric (second fundamental form) relative to the (centroaffine) transverse field $y-u$ (cf.\ \cite{nomizu94}). Namely, let $D$ be the flat affine connection due to the affine structure of $\mathscr{P}_u$, and let $X,Y$ be vector fields on $\mathscr{S}_u$. At point $y\in \mathscr{S}_u$ we can split the derivative $D_XY$ in terms proportional to $y-u$ and tangent to  $\mathscr{S}_u$ according to
\[
D_X Y=\nabla_X Y- h(X,Y) (y-u)
\]
where the former term can be shown to define a covariant derivative. The symmetric tensor $h$ is by definition the {\em affine metric} of $\mathscr{S}_u$. It can be completely deduced from the domain of allowed velocities $D_u$.

Since $Y$ is tangent to $\mathscr{N}_x$, $0=g_y(y,Y)=\dd \mathscr{L}\vert_y(Y)$ and hence
\[
\dd \mathscr{L}(D_X Y)= \!\frac{\p \mathscr{L}}{\p y^\alpha} \left(\!X^\beta \frac{\p Y^\alpha}{\p y^\beta}\!\right)\!= \!X^\beta \frac{\p }{\p y^\beta} \left(\frac{\p \mathscr{L}}{\p y^\alpha} Y^\alpha\!\!\right)-X^\beta Y^\alpha\!  \frac{\p^2 \mathscr{L}}{\p y^\alpha \p y^\beta}\!=-g_y(X,Y),
\]
thus
\[
-g_y(X,Y)=\dd \mathscr{L}(D_X Y)= -h(X,Y) g_y (y, y-u)=  g_y(y,u) h(X,Y).
\]
In conclusion on $\mathscr{S}_u$
\begin{equation}
\gamma= -g_y(y,u) h,
\end{equation}
thus independently of the spacetime dimension
\begin{proposition}
The sky and the affine metric are conformally related. The map between skies of different observers is conformal with respect to the affine metrics.
\end{proposition}
  The specific relationship between $\gamma$ and $h$ requires more data. Having given a Finsler Lagrangian one can calculate $g_y(y,u)$ under the condition that $g_u(u,y)=-1$ and so determine the direction dependent conformal factor. Of course in the Lorentzian case, since the metric is independent of the index, $g_y(y,u)=g_u(u,y)=-1$, so the two metrics coincide.

\section{Conclusions}

We  gave a simple proof of the conformal transformation of night sky. The proof did not use Lorentz transformations and in fact we showed that the result does not depend on the  homogeneity of the indicatrix. The  conformal factor turned out to be a squared Doppler factor. Using this result we  argued that whenever the observers can agree on the photons to be measured, they can also single out a common sky metric just by using the spectral (thermal, color) properties of the images. This invariant metric, by depending on just the spacetime event, can in principle be used for cosmological navigational purposes.

\section*{Acknowledgments}
This work has been partially supported by GNFM of INDAM.\\


\end{document}